\documentclass[journal,10pt]{IEEEtran}

\usepackage{graphicx}
\usepackage [noadjust]{cite} 
\usepackage{bm}  
\usepackage{amsmath}
\usepackage{amssymb}
\usepackage{enumerate}
\usepackage{stfloats}
\usepackage{cases}
\usepackage{epstopdf}
\usepackage{soul,color} 
\usepackage{hyperref}
\usepackage[percent]{overpic}
\usepackage{tabularx}
\usepackage[table]{xcolor}
\usepackage{multirow}
\usepackage{booktabs}  
\usepackage{tabu} 

\usepackage{algorithm}
\usepackage{algpseudocode}%

\newcommand{\alg}[1]{
	\ifnum\value{algcounter}<10{
		\begin{enumerate}[\arabic{algcounter}.\hspace{\numexpr\value{algblockspace}+0 pt}]
			\setlength\itemindent{5pt}
			\item \hspace{-6pt} #1
		\end{enumerate} \stepcounter{algcounter}
	}\else{
		\begin{enumerate}[\arabic{algcounter}.\hspace{\numexpr\value{algblockspace}+0 pt}]
			\setlength\itemindent{0pt}
			\item \hspace{-6pt} #1
		\end{enumerate} \stepcounter{algcounter}
	}
	\fi
}

\newtheorem {remark}{Remark}

\begin{document}
	\raggedbottom
	\allowdisplaybreaks
	\title{A Federated Deep Learning Framework for Cell-Free RSMA Networks}
	\author{S. Ali Mousavi, Mehdi Monemi, \textit{Member}, IEEE, Reza Mohseni, Matti Latva-aho, \textit{Fellow}, IEEE 
		\thanks{
}
\thanks{
  Mehdi Monemi and Matti Latva-aho are with the Centre for Wireless Communications (CWC), University of Oulu, 90570 Oulu, Finland (emails: mehdi.monemi@oulu.fi, matti.latva-aho@oulu.fi).
}
\thanks{
S. Ali Mousavi and Reza Mohseni are with the Department of Electrical Engineering, Shiraz university of Technology, Shiraz 71557-13876, Iran (e-mails: al.mousavi@sutech.ac.ir, mohseni@sutech.ac.ir).
}
	}
	\maketitle
	\begin{abstract}
		
		Next-generation wireless networks are poised to benefit significantly from the integration of three key technologies (KTs): Rate-Splitting Multiple Access (RSMA), cell-free architectures, and federated learning. Each of these technologies offers distinct advantages in terms of security, robustness, and distributed structure. In this paper, we propose a novel cell-free network architecture that incorporates RSMA and employs machine learning techniques within a federated framework. This combination leverages the strengths of each KT, creating a synergistic effect that maximizes the benefits of security, robustness, and distributed structure.
		We formally formulate the access point (AP) selection and precoder design for max-min rate optimization in a cell-free MIMO RSMA network. Our proposed solution scheme involves a three-block procedure. The first block trains deep reinforcement learning (DRL) neural networks to obtain RSMA precoders, assuming full connectivity between APs and user equipments (UEs). The second block uses these precoders and principal component analysis (PCA) to assign APs to UEs by removing a subset of AP-UE connections. The final block fine-tunes the RSMA precoders by incorporating the associated APs into a second DRL network. To leverage the distributed nature of the cell-free network, this process is implemented in a Federated Deep Reinforcement Learning (FDRL) structure operating through the cooperation of APs and a central processing unit (CPU). 
		Simulation results demonstrate that the proposed FDRL approach performs comparably to a benchmark centralized DRL scheme.  
		Our FDRL approach, provides a balanced trade-off, maintaining high performance with enhanced security and reduced processing demands.

	\end{abstract}
	
	
	\thispagestyle{empty}
	
	\section{Introduction}
	Cell-free communication is a promising technique that offers significant improvements in user coverage and spectral efficiency compared to traditional cellular communication. This approach involves establishing a network of distributed access points (APs), connected to a central processing unit (CPU) for data transmission with user equipments (UEs) \cite{CF1}-\cite{channel}. Additionally, it enhances the diversity of channels between APs and UEs. The performance of cell-free communication is directly influenced by AP selection and the UE clustering scheme. Furthermore, incorporating multiple-input multiple-output (MIMO) technology can additionally enhance cell-free communication. By optimizing the precoding matrices used in MIMO cell-free systems, overall performance can be greatly improved. However, updating precoding matrices and handling computational processes might impose extreme processing loads in large dimensional cell-free networks.	
	Performing processing, control, and signaling tasks in the CPU optimizes network performance but puts a heavier computational load on the CPU, resulting in increased delay. To address this issue, a decentralized approach can be adopted where the processing and computation of network parameters are distributed among different nodes \cite{FL-CF1}, \cite{FL-CF2}. Decentralized methods generally reach sub-optimal solutions (compared with the potentially optimal solutions of centralized schemes). Besides, they involve back-and-forth communication between APs and the CPU. However, they offer several advantages; Firstly, they significantly reduce delay compared to the centralized methods by reducing processing load. Additionally, for applications that require high data security, this method ensures better security guarantees. This is mainly due to the distributed nature of these methods, which limits the aggregation and sharing of network information. Federated learning is regarded as one such distributed machine learning scheme implemented for resource allocation in wireless networks \cite{FL-CF3,FL-CF4,FL-CF5}.
	The structure of cell-free networks is well-aligned with the distributed nature of the federated learning framework. In cell-free networks, the tasks traditionally handled solely by the CPU can be off-loaded across each AP using a federated learning framework. Employing federated learning in cell-free networks ensures that each UE's data is localized for the relevant AP, harnessing the processing power of each AP while simultaneously reducing the backbone processing and overall costs associated with cell-free networks. 
	
	
	
	%
	
	To further benefit from the pioneering multiple access technologies, we integrate the rate-splitting multiple access (RSMA) into multi-user cell-free communications leveraging federated leaning framework.  
	In contrast to previous generations of multiple access techniques such as orthogonal/non-orthogonal multiple access (OMA/NOMA), and space division multiple access (SDMA), RSMA brings numerous advantages, including a higher level of security and robustness, as well as spectral, energy, and computational efficiency \cite{RSMA1}, \cite{RSMA2}. The RSMA technique demonstrates remarkable flexibility in handling different interference levels, network loads, services, traffic patterns, and UE deployments. It mitigates the necessity of accurate channel state information (CSI), leading to being resilient against mixed-critical quality-of-service demands \cite{RSMA-CF1}-\cite{RSMA-CF7}. 
	RSMA encompasses a broad range of multi-user strategies that are based on the rate-splitting (RS) principle. In RS, a UE message is divided into multiple parts which can be decoded flexibly by one or more receivers. 
	The message of each UE is split into a common and a private part. The common parts of all UEs' messages are aggregated to form a collective message that can be decoded by all UEs. The private part of each UE message is individually encoded in the transmitter, and then decoded at the receiver by the corresponding UE independent of others. For each UE, the common message is concatenated with the corresponding private message and then transmitted toward the corresponding UE. By tuning the power allocated to each of the common and private parts of the message, RSMA facilitates enhanced interference reduction. 
	
	This paper introduces a novel federated learning-based framework for optimizing resource allocation in cell-free RSMA networks. Our proposed approach addresses the challenges of centralized control and data privacy by enabling multi-level distributed learning across multiple APs. By jointly optimizing AP selection and precoder design in a federated scheme, our framework aims to maximize RSMA network performance in terms of max-min fairness criteria.
	
	
	\subsection{Background Work}
	Numerous studies have extensively explored various aspects of traditional cell-free communication, highlighting both the inherent opportunities and associated challenges to next-generation wireless technology \cite{CF1}-\cite{channel}. Recent research has also focused on integrating federated learning within cell-free communications \cite{FL-CF1}-\cite{FL-CF5}. In particular, the authors of \cite{FL-CF1} showcased the potential of cell-free massive MIMO to enable federated learning, thereby preserving data security. Additionally, in \cite{FL-CF3} a federated multi-agent reinforcement learning algorithm is employed for AP selection and clustering, resulting in enhanced performance of the system for dynamic network environments. Furthermore, the work presented in \cite{FL-CF4} introduced horizontal and vertical decentralized precoding methods using the federated learning framework. The proposed scheme effectively reduced computational complexity and communication overhead, thereby minimizing latency in practical network applications.
	To implement federated learning in wireless network problems, deep reinforcement learning (DRL) techniques such as deep deterministic policy gradients (DDPG) have proven to be effective in tackling complex problems like channel estimation, signal detection, power allocation, and precoding matrix computation in large-scale dimensions  \cite{Fl-cf-DRL1}-\cite{Fl-cf-DRL4}. The integration of DRL with federated learning demonstrates significant efficiency in cell-free networks, enabling each AP to locally contribute toward a global objective leveraging DRL benefits \cite{FL-CF3}.
	
	From a multiple access technology point of view, RSMA has been investigated in many works as one of the key enabling components for next-generation wireless networks. A comprehensive study examining the foundational principles and various aspects of the RSMA has been explored in \cite{RSMA1,RSMA2}. Moreover, the works \cite{RSMA-CF1,RSMA-CF2,RSMA-CF3-SR1,RSMA-CF5-MM,RSMA-CF6} provide an in-depth analysis of RSMA applications within cell-free networks. Among these, the authors of \cite{RSMA-CF1} introduce innovative cluster-based linear precoders designed for cell-free networks, considering the challenge of imperfect CSI. The application of RSMA to massive machine-type communications as well as simultaneous wireless information and power transfer has been investigated in \cite{RSMA-CF2} and \cite{RSMA-CF3-SR1} respectively.
	The authors of \cite{RSMA-CF5-MM} focus on finding the optimal beamformer and RSMA rate coefficients by addressing the max-min rate optimization problem. The authors of \cite{RSMA-CF6} have proposed a resource allocation algorithm for ultra-reliable and ultra-low latency (URLLC) communications aiming to optimize the energy efficiency in RSMA cell-free Multiple-Input Single-Output (MISO) networks, considering transceiver impairments. 
	
	The AP selection is an important part of resource allocation investigated in many cell-free MIMO networks
	\cite{APselection}-\cite{APselection2}. When dealing with RSMA cell-free networks, the optimal AP selection becomes a more complicated issue due to the complex and highly coupled relation between the optimal decision variables including complex-valued precoders and binary-valued UE-AP associations. 
	In this regard, most RSMA cell-free networks either ignore the AP selection subproblem by considering a given pre-allocation of the APs to the UEs \cite{RSMA-CF2}-\cite{RSMA-CF6}, or try to decouple the AP selection from the allocation of other resources, proposing a scheme to solve the former based on a simplistic analysis of channel gains \cite{RSMA-CF7}. In this work, we integrate a more efficient DRL-based AP selection into precoder design for RSMA cell-free networks.


	\begin{figure}
		\centering
		\includegraphics [width=230pt,height=160pt]{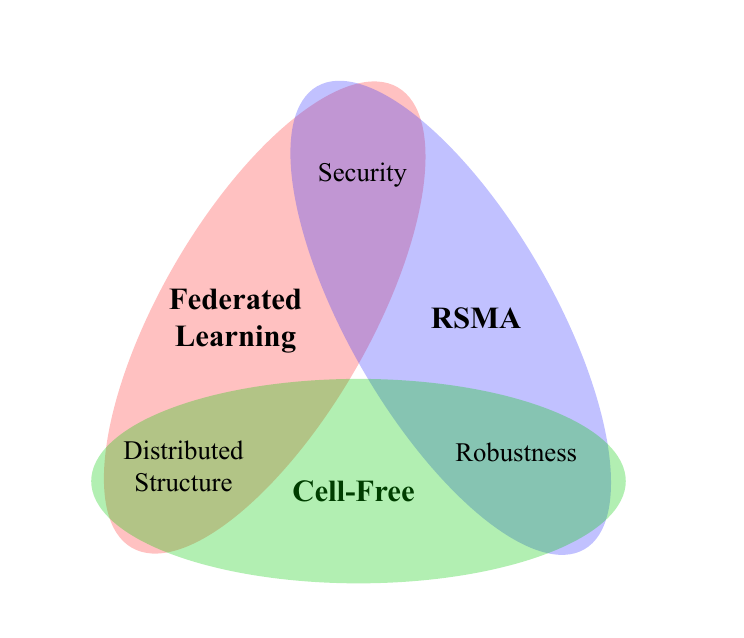} \\
		\caption{The relation between the KTs and KPs incorporated in our proposed structure.} 
		\label{fig:combine}
	\end{figure}
	
	\subsection{Motivation and Contributions}
	The main contributions of this work are expressed as follows:	
	\begin{itemize}
		\item  
		Next-generation wireless networks leverage three key technologies (KTs): RSMA, cell-free, and federated learning. These KTs offer three key properties (KPs): security, robustness, and distributed structure. As depicted in Fig. \ref{fig:combine}, each KT is associated with two KPs and each KP is shared by two KTs. This interconnectedness suggests that combining these KTs can leverage the benefits of all KPs. In this work, we propose a novel cell-free network architecture that employs RSMA in the physical layer and utilizes machine learning techniques in a federated scheme. This framework leverages the strengths of each KT, creating a synergistic effect. 
		
		\item We formally formulate the AP selection and precoder design for the max-min rate optimization in a cell-free RSMA network. We propose a solution scheme based on federated learning in a procedure involving 3 blocks. The first block neglects the AP selection and trains the corresponding DRL neural networks (NNs) to obtain RSMA precoders. In the second block, by using the precoders obtained from the first one and applying the principle component analysis (PCA), we address the assignment of APs to the UEs in the cell-free network environment. Finally, in the last block, we employ the incorporation of the associated APs as the inputs to a second DRL network to fine-tune the RSMA precoders. The overall solution scheme is proposed as a distributed learning structure termed {\it federated deep reinforcement learning (FDRL)} wherein the stated blocks operate in a federated scheme through the cooperation of the agents (i.e.,APs) and CPU.
		
		\item The performance of the proposed solution scheme is evaluated through simulations and compared to a centralized DRL scheme, which serves as a benchmark. Although the centralized DRL approach might generally exhibit rather superior performance compared to the sub-optimal FDRL approach, this advantage comes at the cost of increased processing load and reduced security. Our simulations demonstrate that the performance of the proposed FDRL approach is closely comparable to that of the centralized DRL approach.
		
		
	\end{itemize}
	
	
	The remaining sections of this paper are organized as follows. Section II presents the system model and problem formulation. Section III elaborates on the proposed solution scheme. Numerical results and discussion are presented in Section IV. Finally, the paper is concluded in Section V.
	
	\textit{Notation}: Italic letters like $ a $ or $ A $ denote scalars, whereas boldface uppercase $ (\bold{A}) $ and lowercase $ (\bold{a}) $ letters denote matrices and vectors, respectively. Calligraphic variables like $\mathcal{A}$ represent sets, with $ |\mathcal{A}| $ being the cardinality. $ \mathrm{Mat2Vec} ( \bold{A}) $ is a function that maps the matrix $  \bold{A}_{MN} $ to a vector of dimension $ MN\times 1 $. $ \bold{A}^{\mathrm{H}} $ denotes the Hermitian (conjugate) transpose and $\|\bold{A}\|_F$ is the Frobenius norm of matrix $ \bold{A} $.  
	The determinant of $ \bold{A} $ is denoted by $ det (\bold{A}) $. $ \bold{I}_n $ is the $ n \times n $ identity matrix. 
	
	\section{System Model and Problem Formulation}
	\subsection{System Model}	
	\label{sec:system_model}
\begin{figure}
		\centering
		\includegraphics [width=254pt]{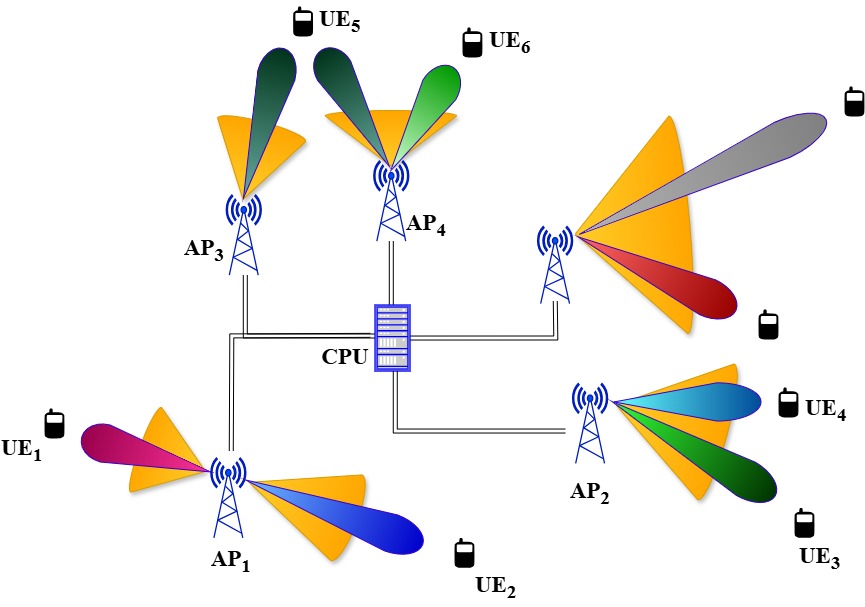} \\
		\caption{The system model of a cell-free MIMO network including APs, UEs, CPU and fronthaul connections. The APs transmit both common and private signal parts using the RSMA, illustrated by wide yellow beams for common signals and narrow beams for private signals. 
		} 
		\label{fig:system}
	\end{figure}

	We study a downlink cell-free MIMO network as shown in Fig.~\ref{fig:system} where a set of $K$ UEs denoted by $\mathcal{K}=\{1,2,..., K\}$ are served by a set of $N$ APs denoted by $\mathcal{N}=\{1,2,.., N\}$. The APs are coordinated by a CPU. Each AP and UE is equipped with $ M $ and  $ M' $ antennas respectively.   
	We assume that UEs are randomly distributed in the network and each UE has at least one serving AP in its vicinity. Each AP $ n \in \mathcal{N} $ serves a set of UEs $ \mathcal{K}_n \subset \mathcal{K}$. The parameter $ g_{kn}\in \{0,1\} $  denotes a binary indicator related to AP selection, representing whether UE $ k $ is associated with AP $ n $. 
	The signal transmitted from AP $n$ is represented by $\bold{x}_n\in \mathbb{C}^{M \times 1} $ and  the channel matrix between UE $k$ and AP $n$ is denoted by $\bold{H}_{kn}\in \mathbb{C}^{M \times M'} $. The signal received by UE $k$, represented by $\bold{y}_k\in \mathbb{C}^{M' \times 1}$, is obtained as follows:
	\begin{align}
		\bold{y}_k= \sum_{n\in \mathcal{N}} \bold{H}_{kn}^{\mathrm{H}} \bold{x}_n + \bold{\nu}_k
		\label{eq1} 
	\end{align}
	where $ \bold{\nu}_k\sim \mathcal{C}\mathcal{N}(0,N_0\bold{I}_{M'})$ represents the noise in which $N_0= K'TBN^F$ is the noise power, $K'$ is Boltzmann constant, $T$ is temperature and $N^\mathrm{F}$ is noise figure. The channel elements are modeled as $[\bold{H}_{kn}]_{ij}=\xi_{ij}\sqrt{\kappa_{kn}}$ where $\xi_{ij}$ represents the small-scale fading which follows independent but not identically distributed (i.n.i.d) Nakagami distribution with spreading and shape parameters $ \mathcal{M} $ and $ \Omega $, respectively. 
	Furthermore $\kappa_{kn}$ represents the large-scale fading which is modeled as $\kappa_{kn} = PL_{kn}. 10^{\frac{\sigma^{\mathrm{sh}}.z_{kn}}{10}}$ where $PL_{kn}$ is the path-loss obtained as
	\begin{align}
		PL_{kn}&=
		\begin{cases}
			-L-35\log_{10}{d_{kn}} & d_{kn}\ge \overline{d}\\
			-L-15\log_{10}{\overline{d}}-20\log_{10}{d_{kn}} &  \underline{d}<d_{kn}<\overline{d}\\
			-L-15\log_{10}{\overline{d}}-20\log_{10}{\underline{d}} &  d_{kn}\le \underline{d}
		\end{cases}
		\label{eq2}
	\end{align}
	and $10^{\frac{\sigma^{\mathrm{sh}}.z_{kn}}{10}}$ represents the shadow fading with standard deviation $\sigma^{\mathrm{sh}}$ equal to 8dB and $z_{kn}= \sqrt{\epsilon} b_{kn} +  \sqrt{1- \epsilon} q_{kn} $ where $ b_{kn} , q_{kn} \sim \mathcal{N}(0,1)$ and $0\le \epsilon \le 1$ \cite{channel}. In \eqref{eq2}, $d_{kn}$ indicates the distance between UE $k$ and AP $n$, $\overline{d}$ and $\underline{d}$ are fixed given distances and
	\begin{align}
		L =&\ 46.3\ +\ 33.9\log_{10}{f}\ -\ 13.82\log_{10}{h^{\mathrm{AP}}}\notag\\
		&-(1.1\log_{10}{f}\ -0.7)\ h^{\mathrm{UE}}
		+1.56\log_{10}{f}-0.8 
		\label{eq3} 	
	\end{align}
	In \eqref{eq3}, $h^{\mathrm{AP}} $ and $h^{\mathrm{UE}} $ are APs' and UEs' height, and $f$ is carrier frequency in MHz.
	We define $ \boldsymbol{W}_k $  as the message intended for UE $k$, which consists of a common component $ \boldsymbol{W}^\mathrm{c}_k $ and a private component $ \boldsymbol{W}^\mathrm{p}_k $. The common components of all UEs are aggregated to form a single common stream $ \boldsymbol{W}^\mathrm{c} =[\boldsymbol{W}^\mathrm{c}_1, . . . ,\boldsymbol{W}^\mathrm{c}_K]$. Utilizing dedicated common and private codebooks, $\boldsymbol{W}^\mathrm{c}$ and $  \boldsymbol{W}^\mathrm{p}_k $ are encoded into a common stream $ \bold{x}^\mathrm{c}$ and private streams $ \bold{x}^\mathrm{p}_k$, respectively. Notice that the common stream is meant to be decoded by all UEs but not necessarily intended to all of them. On the other hand, private components are specifically designed to be decoded by the corresponding UEs only. For example, it is seen from Fig. \ref{fig:system} that in the proposed cell-free RSMA structure, UE$_5$ receives the related common and private signals from both AP$_3$ and AP$_4$, while UE$_6$ receives the corresponding signals only from AP$_4$.
	Based on the RSMA method, the signal transmitted from AP $n$ is obtained by the combination of common and private parts formulated as follows \cite{RSMA1}, \cite{RSMA-CF2}: 
	\begin{align}
		\bold{x}_n=\bold{P}_n^{\mathrm{c}} \bold{x}^{\mathrm{c}} + \sum_{k\in \mathcal{K}_n} \bold{P}_{kn}^{\mathrm{p}} \bold{x}_k^{\mathrm{p}} 
		\label{eq4} 
	\end{align}
	where $ \bold{P}_n^{\mathrm{c}}\in \mathbb{C}^{M \times M'} $ and $  \bold{P}_{kn}^{\mathrm{p}} \in \mathbb{C}^{M \times M'} $ are  common and private precoders respectively. Also, $ \bold{x}^\mathrm{c} \in \mathbb{C}^{M' \times 1} $ and $ \bold{x}_k^\mathrm{p} \in \mathbb{C}^{M' \times 1}$ are respectively the common and private part of the signal of UE $ k $ where $\mathbb{E}\{\|\bold{x}_k^\mathrm{p}\|^2\}=\mathbb{E}\{\|\bold{x}^\mathrm{c}\|^2\}=1,  \forall k\in \mathcal{K}$. 
	
	Based on successive interference cancellation and inspired by \cite{RSMA1}, the rate of the common and private part of the RSMA method in the cell-free network is respectively expressed as follows:
	\begin{subequations}
		\label{eq5} 
		\begin{align}
			R_k^{\mathrm{c}}
			&=
			\log_2{\mathrm{det}(\bold{I}_{M'} + \sum_{n\in \mathcal{N}}\boldsymbol{\Xi}_{kn}^{\mathrm{c}}}) 
			\label{eq51} 
			\\
			R_k^{\mathrm{p}}
			&=
			\log_2{\mathrm{det}(\bold{I}_{M'} + \sum_{n\in \mathcal{N}}{g_{kn}\boldsymbol{\Xi}^{\mathrm{p}}_{kn})}} 
			\label{eq52} 
		\end{align}
	\end{subequations}
	where 
	\begin{subequations}
		\label{eq500} 
		\begin{align}
			\boldsymbol{\Xi}^{\mathrm{c}}_{kn}
			&=
			({\bold{P}_n^{\mathrm{c}}})^{\mathrm{H}} \bold{H}_{kn}(\boldsymbol{\Sigma}_k^{\mathrm{c}})^{-1}(\bold{H}_{kn})^{\mathrm{H}} \bold{P}_n^{\mathrm{c}}
        \label{eq501} 
        \\
        \boldsymbol{\Xi}^{\mathrm{p}}_{kn}
        &=
	(\bold{P}_{kn}^{\mathrm{p}})^{\mathrm{H}} \bold{H}_{kn}(\boldsymbol{\Sigma}_k^{\mathrm{p}})^{-1}(\bold{H}_{kn})^{\mathrm{H}} \bold{P}_{kn}^{\mathrm{p}}
			\label{eq502} 
		\end{align}
	\end{subequations}
	In \eqref{eq5}, $R_k^{\mathrm{c}}$ and $R_k^{\mathrm{p}}$ represent the common and private parts of the transmission rate of UE $k$. The matrices $\boldsymbol{\Sigma}_k^{\mathrm{c}}$ and $\boldsymbol{\Sigma}_k^{\mathrm{p}}$ in \eqref{eq500}, are the common and private interference terms respectively, obtained as follows:
	\begin{align}
		\boldsymbol{\Sigma}_k^{\mathrm{c}}=N_0 \bold{I}_{M'} + \sum_{n\in \mathcal{N}}\sum_{i\in \mathcal{K}} \bold{H}_{kn}^{\mathrm{H}} \bold{P}_{in}^{\mathrm{p}} {\bold{P}_{in}^{\mathrm{p}}}^{\mathrm{H}} \bold{H}_{kn}  \notag\\
		\boldsymbol{\Sigma}_k^{\mathrm{p}}=N_0 \bold{I}_{M'} + \sum_{n\in \mathcal{N}}\sum_{i\in \mathcal{K}, i \ne k} \bold{H}_{kn}^{\mathrm{H}} \bold{P}_{in}^{\mathrm{p}} {\bold{P}_{in}^{\mathrm{p}}}^{\mathrm{H}} \bold{H}_{kn} 
		\label{eq6} 
	\end{align}
	The total achievable common rate $R^{\mathrm{c}}$ is calculated as 
	\begin{align}
		R^{\mathrm{c}}&= \underset{k}{\text{min}}\ \ R_k^{\mathrm{c}}
		\label{eq7} 
	\end{align}
	Let $ c_k R^{\mathrm{c}}$ denoted to the portion of the common rate allocated to UE $ k $. To successfully decode the common stream $\bold{x}^{\mathrm{c}} $, the rate of the common part should satisfy the $\sum_{k\in \mathcal{K}}c_k R^{\mathrm{c}} \le R^c$. This results in
	\begin{align}
		\label{eq:7367543}
		\sum_{k\in\mathcal{K}} c_k\le 1
	\end{align}
	The total rate of UE $  k $ is then expressed as follows:
	\begin{align}
		R_k= c_k R^{\mathrm{c}} + R_k^{\mathrm{p}}
		\label{eq8} 
	\end{align}
	\subsection{Problem Formulation}
	Based on what has been stated so far, given the set of UEs and APs in a cell-free RSMA network, we aim to maximize the minimum rate while considering all related constraints. This is formally defined as follows:   
	\begin{align} \mathcal{P}1:	&\ \underset{g_{kn}, c_k,  \bold{P}_{n}^{\mathrm{c}}, \bold{P}_{kn}^{\mathrm{p}}}{\text{max \ min}}\  R_k\notag\\
		\mathrm{s.t.}\ 
		&\ 
		C1:\|\bold{P}^{\mathrm{c}}_n\|^2_F+\sum_{k\in \mathcal{K}} g_{kn} \|\bold{P}_{kn}^{\mathrm{p}}\|^2_F\le P^{\mathrm{max}}_n,\ \ \forall n
		\notag\\ 
		&\ C2: \sum_{n\in \mathcal{N}}g_{kn} \le \overline{N}^{\mathrm{UE}},\ \ \forall k \notag\\
		&\ C3: \sum_{k\in \mathcal{K}}c_k \le 1,\ \ \forall n\notag\\
		&\ C4: g_{kn} \in \{0,1\},\ \ \ \forall k,n \notag\\
		&\ C5: c_k\ge 0,\ \ \ \forall k
		\label{eq9} 
	\end{align}
	where C1 guarantees that the total transmit power of AP $n$ be lower than the threshold value $P^{\mathrm{max}}_n$, 
	C3 is the common rate portion allocation constraint corresponding to \eqref{eq:7367543}, C4 identifies the association of AP to the UEs, and finally, C5 ensures that the allocated portions of the common rate are non-negative. 
	Problem $ \mathcal{P}1 $ is a mixed-integer non-linear program with highly coupled decision variables rendering it a complex non-convex problem. 
	
	\section{Proposed Solution Scheme}
	In this section, we outline a comprehensive strategy for addressing Problem $ \mathcal{P}1 $. We approach the solution by decomposing $ \mathcal{P}1 $ into two sub-problems.  First, in subsection A, we present a solution scheme based on PCA and integer linear programming to determine the parameters $ g_{kn}, \forall k,n$. Next, in subsection B, we propose a federated DRL structure to determine the parameters $c_k,\ \bold{P}_n^{\mathrm{c}},\  \bold{P}_{kn}^{\mathrm{p}}$ under the assumption that the AP selection has been completed and the parameters $ g_{kn}, \forall k,n$ have been determined. The reward computation associated with the FDRL algorithm is discussed in subsection C. Subsequently, subsection D introduces the overall algorithm designed to solve the main problem presented in this paper, detailing a three-step solution process. Finally, we analyze the computational complexity in the last subsection.
	
	\subsection{AP Selection}
	To facilitate the selection of the AP, let the precoding matrices and allocation of the common rate portion be available according to the mechanism that will be expressed in Section \ref{sec:FDRL}. Problem $ \mathcal{P}1 $ is then reduced as follows:
	\begin{align} \mathcal{P}2:	&\ \underset{g_{kn}}{\text{max \ min}}\  R_k\notag\\
		\mathrm{s.t.}\ 
		&\ C1,\ C2,\ C4
		\label{eq93} 
	\end{align}
	In $ \mathcal{P}2 $, all constraints are linear. To transfer $ \mathcal{P}2 $ into a linear program, we need to linearize the objective function.
	As observed in \eqref{eq5} and \eqref{eq8}, the objective function of $ \mathcal{P}2 $ is influenced by $ g_{kn} $ only through the private rate component of the UEs. Hence, we may equivalently write $ \mathcal{P}2 $ as follows:
	\begin{align} 
		\underset{g_{kn}}{\mathrm{max\ \ \!\! \!  min}} & \  \mathrm{det}(\bold{I}_{M'} + \sum_{n\in \mathcal{N}}{g_{kn}\boldsymbol{\Xi}^{\mathrm{p}}_{kn}})\notag
		\\
		\mathrm{s.t.}\ 
		&\ C1,\ C2,\ C4
		\label{eq94} 
	\end{align}
	In the case of single-antenna UEs (i.e., MISO) where $M'=1$, the objective function of $\eqref{eq94}$ turns into a linear function of $g_{kn}$. For the general MIMO case where $M'>1$, we employ a dimensionality reduction procedure using the PCA method. This method is utilized for simplifying complexity and reducing dimensionality for MIMO communication \cite{PCA1}, \cite{PCA2}. Here, we transform the AP selection problem for the multiple antenna UE to an equivalent approximated problem wherein, for the private part of the rate of each UE, we consider the dominant contributor of $\boldsymbol{\Xi}_{kn}^{\mathrm{p}}$ corresponding to the largest eigenvalue of this matrix. This approximates the complex representation of \eqref{eq94} into the following linear integer program:
	\begin{align} \mathcal{P}3:	&\ \underset{g_{kn}}{\text{max \ min}}\  \sum_{n \in \mathcal{N}}g_{kn} \lambda^{max}_{kn}\notag\\
		\mathrm{s.t.}\ 
		&\ C1,\ C2,\ C4
		\label{eq96} 
	\end{align}
	where $\lambda^{max}_{kn}$ is maximum eigenvalue of $\boldsymbol{\Xi}_{kn}^{\mathrm{p}}$.
	
	The AP selection problem $ \mathcal{P}3 $ is an integer linear program that can be solved using standard solvers such as CVX and Mosek. Note that in case we have $M'=1$, $\mathcal{P}2$ and $\mathcal{P}3$ become equivalent, resulting in the globally optimal solution of the AP assignment. 
	
	\subsection{Federated Deep Reinforcement Learning}
	\label{sec:FDRL}
	Given $ g_{kn} $ calculated from the previous subsection, $ \mathcal{P}1 $ is written as the following non-convex problem:
	\begin{align} \mathcal{P}4:	&\ \underset{c_k,  \bold{P}_{n}^{\mathrm{c}}, \bold{P}_{kn}^{\mathrm{p}}}{\text{max \ min}}\  R_k\notag\\
		\mathrm{s.t.}\ 
		&\ \|\bold{P}^{\mathrm{c}}_n\|^2_F+\sum_{k\in \mathcal{K}_n} \|\bold{P}_{kn}^{\mathrm{p}}\|^2_F\le P^{\mathrm{max}}_n,\ \ \forall n
		\notag\\ 
		&\ C3,\ C5
		\label{eq91} 
	\end{align}
	 To cope with the limitations of existing solution approaches and leverage the benefits of the distributed nature of cell-free networks, we utilize a DRL-based solution approach by implementing the corresponding computations within a federated learning framework. To elaborate more, in what follows we discuss both elements associated with the proposed FDRL framework, namely the {\it DRL} and {\it federated learning} structures.
	
\begin{figure*}[t]
		\centering
		\includegraphics[height=250pt,width=350pt]{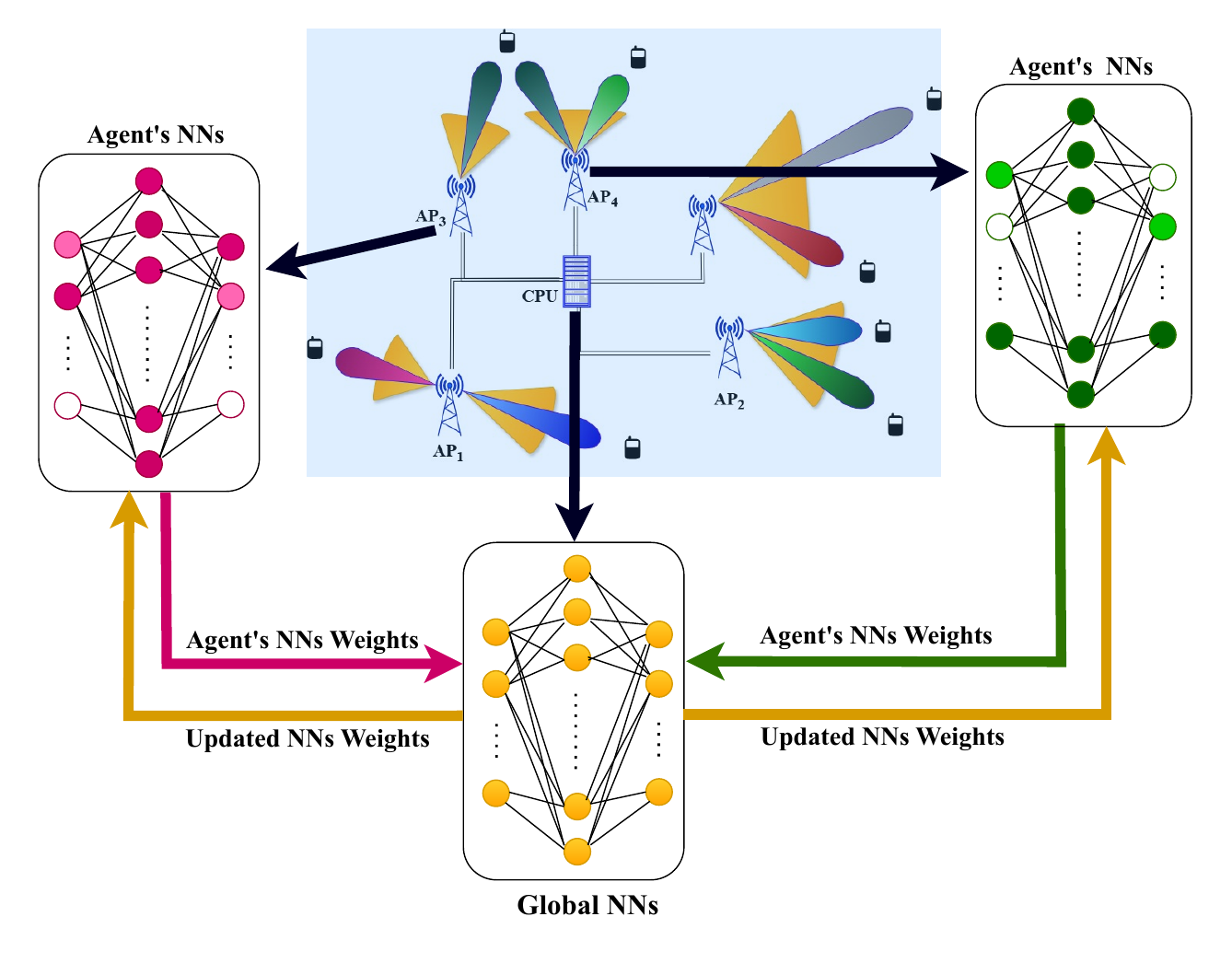}
		\hspace{-10pt}
		\caption{FDRL process in the cell-free RSMA network. Each agent $n$ shares its NN's weights $\boldsymbol{\theta}_n$ with the CPU. After the aggregation of all agents' parameters, the updated weights of the global NN are transmitted back to the agents. }
		\label{fig:FDRL}
	\end{figure*}

	DRL is a powerful framework that combines reinforcement learning principles with deep learning techniques, enabling agents to learn optimal policies through interaction with complex environments. In scenarios where both the action and state spaces are continuous, the DDPG algorithm becomes particularly advantageous, utilizing an actor-critic architecture that allows for the approximation of both the policy and value functions. To reduce the CPU load, we propose localized precoding computation, where each agent has access only to its own CSI for the UEs it serves.  Therefore, there is no need to exchange CSI from other APs over the fronthaul, and only the NNs' parameters need to be exchanged between CPU and agents at some iterations. Noting that the local CSI is employed in the FDRL framework, each agent is unable to have real-time coordination with other agents to optimize the calculated precoding vector. This means that the proposed scheme is sub-optimal in nature; while achieving closely competitive results with the centralized solution, it enhances data security and reduces computational load through distributed processing.
	Moreover, we limit the interaction between the CPU and the APs to periodically share the trained information as illustrated in Fig. \ref{fig:FDRL}. Here we assume that each AP operates as an agent having its own local state space and action space. While these states and actions may vary across agents, they all have identical dimensions, as explained later in this subsection. Furthermore, all agents employ identical NNs structures. 
	
	We present the FDRL as a Markov decision process, which is defined as a set of tuples denoted by $ (\mathcal{S},\mathcal{A},r,\gamma) $ where each of the agents as well as the CPU are assigned with an instance of the tuple. Here, $ \mathcal{S} $ represents the state space, $ \mathcal{A} $ represents the action space, $ r $ corresponds to the reward function, and $\gamma$ refers to the discount factor. 
	Let $T'$  be the total number of iterations each agent updates its neural network training parameters locally before exchanging them with the CPU. The FDRL updating process consists of two types of iterations. For iterations $t\neq mT', m\in\mathbb{N_+}$, each agent updates its DRL policy locally. For iterations $t=mT$, all agents' learnable parameters are sent to the CPU and combined at the CPU; the updated parameters are then sent back to the agents.
	At each iteration $t\neq mT'$, each agent $n$ observes a state vector $ \boldsymbol{s}_n(t) \in \mathcal{S} $ corresponding to a set of related environment parameters based on which, the agent decides to select the action vector $\boldsymbol{a}_n(t) \in \mathcal{A}$ according to the policy $\pi_n$. 
    
	Let's define $t_0=mT'$.  
	For each iteration $ t_0 < t < t_0+ T'$, the state and action vectors of agent $n$ are respectively formulated in a federated scheme as follows:
	\begin{align}
		\boldsymbol{s}_n(t)
		=&\left[
		\tilde{\boldsymbol{s}}_1(t_0), ..., \tilde{\boldsymbol{s}}_{n-1}(t_0), \tilde{\boldsymbol{s}}_{n}(t), \tilde{\boldsymbol{s}}_{n+1}(t_0),...,, \tilde{\boldsymbol{s}}_{N}(t_0) \right] \notag
		\\
		\boldsymbol{a}_n(t)
		=&\left[
		\tilde{\boldsymbol{a}}_1(t_0), ..., \tilde{\boldsymbol{a}}_{n-1}(t_0), \tilde{\boldsymbol{a}}_{n}(t), \tilde{\boldsymbol{a}}_{n+1}(t_0),...,, \tilde{\boldsymbol{a}}_{N}(t_0)\right]
		\label{eq9911} 
	\end{align}
	
	It is seen that for each iteration $ t_0 \leq t < t_0+ T'$,	while all elements $\tilde{\boldsymbol{s}}_{n'}(.)$ and $\tilde{\boldsymbol{a}}_{n'}(.), \forall n'$ are incorporated in updating agent parameters (including NNs' weights and reward function), the only directly related portion $\tilde{\boldsymbol{s}}_n(t)$ and $\tilde{\boldsymbol{a}}_{n}(t)$ are respectively updated locally; 
	other elements remain fixed as $\tilde{\boldsymbol{a}}_{n'}(t_0)$ and $\tilde{\boldsymbol{s}}_{n'}(t_0), \forall n'\neq n$. 
	Now we deal with defining the components of $\boldsymbol{s}_n(.)$ and $\boldsymbol{a}_n(.)$ related to our RSMA cell-free network model.
	Let's define respectively the received common and private signal parts from AP $n$ at UE $k$ as follows:
	\begin{align}
		\mathrm{\hat{\bold{Y}}}_{kn}^{\mathrm{c}} =& \bold{H}_{kn}^{\mathrm{H}}\bold{P}_n^{\mathrm{c}} \notag \\
		\mathrm{\hat{\bold{Y}}}_{kn}^{\mathrm{p}} =& \bold{H}_{kn}^{\mathrm{H}}\bold{P}_{kn}^{\mathrm{p}}
		\label{eq991} 
	\end{align}	
	For each agent $n$, we consider the elements of $\mathrm{\hat{\bold{Y}}}_{kn}^{\mathrm{c}}$ and $\mathrm{\hat{\bold{Y}}}_{kn}^{\mathrm{p}}$ as the state vector, and the  decision variables of \eqref{eq91} as the action vector for the corresponding agent, presented for each $t$ as follows: 
	\begin{align}
		\tilde{\boldsymbol{s}}_n=& [
		\mathrm{\hat{\bold{y}}}_{1n}^{\mathrm{c}},
		...,
		\mathrm{\hat{\bold{y}}}_{Kn}^{\mathrm{c}},   
		\mathrm{\hat{\bold{y}}}_{1n}^{\mathrm{p}},
		...,
		\mathrm{\hat{\bold{y}}}_{Kn}^{\mathrm{p}}
		]\notag \\
		\tilde{\boldsymbol{a}}_n=& [\bold{p}_{n}^{\mathrm{c}}, \bold{p}_{1n}^{\mathrm{p}},..., \bold{p}_{Kn}^{\mathrm{p}}, c_1,..., c_K]
		\label{eq9910} 
	\end{align}
	
	
	where
	\begin{align}
		\mathrm{\hat{\bold{y}}}_{kn}^{\mathrm{c}}=&\mathrm{Mat2Vec}(\mathrm{\hat{\bold{Y}}}_{kn}^{\mathrm{c}})\notag\\
		\mathrm{\hat{\bold{y}}}_{kn}^{\mathrm{p}}=&\mathrm{Mat2Vec}(\mathrm{\hat{\bold{Y}}}_{kn}^{\mathrm{p}})\notag\\
		\mathrm{\bold{p}}_{n}^{\mathrm{c}}=&\mathrm{Mat2Vec}(\mathrm{\bold{P}}_{n}^{\mathrm{c}})\notag\\
		\mathrm{\bold{p}}_{kn}^{\mathrm{p}}=&\mathrm{Mat2Vec}(\mathrm{\bold{P}}_{kn}^{\mathrm{p}})
	\end{align}
	
	Note that all elements of the action vector are continuous variables. This justifies the employment of the DDPG learning algorithm, as it is specifically designed to handle optimization problems having continuous action space with large dimensions.
	We consider the reward function equivalent to the achieved rate of each agent whose detailed derivation will be presented in the next subsection. Let $ \boldsymbol{\theta}_n =\{\boldsymbol{\theta}_n ^{\mu},\boldsymbol{\theta}_n ^{\mu'}, \boldsymbol{\theta}_n^Q,\boldsymbol{\theta}^{Q'}_{ n }\} $ be the network parameters of agent $n$, where $\boldsymbol{\theta}_n^{\mu}$ and $\boldsymbol{\theta}_n^Q$ are respectively the actor and critic NN weights and $\boldsymbol{\theta}_n^{\mu'}$ and $\boldsymbol{\theta}_n^{Q'}$ are those respectively corresponding to the target actor and target critic NNs. In detail, the DDPG structure consists of the following networks: The actor-network $ \mu_n(\boldsymbol{s}_n|\boldsymbol{\theta}_n ^{\mu})$ which translates the state vector into the action vector, the critic network $Q_n(\boldsymbol{s}_n,\boldsymbol{a}_n|\boldsymbol{\theta}_n^Q)$ that maps the state, as well as the output action of $\mu_n(\boldsymbol{s}_n|\boldsymbol{\theta}_n^{\mu})$ to a Q-value metric, and finally the target actor and critic networks denoted respectively by $\boldsymbol{\theta}_n^{\mu'} $ and $\boldsymbol{\theta}_n^{Q'}$ which are the delayed versions of actor and critic networks.   
	
	
	At each iteration $t\neq m T'$, each agent $n$ takes action $ \boldsymbol{a}_n(t)=\mu_n(\boldsymbol{s}_n(t)|\boldsymbol{\theta}_n^{\mu})+\nu^a$, where $\nu^a$ is an exploration noise, and gets the state $\boldsymbol{s}_n(t+1)$ and calculates the reward  $r_n(t)$. The experience $(\boldsymbol{s}_n(t),\boldsymbol{a}_n(t),r_n(t),\boldsymbol{s}_n(t+1))$ is then stored and added to the experience buffer $\boldsymbol{\mathrm{R}}_n$. A random minibatch $(\boldsymbol{s}_n(\tilde{t}),\boldsymbol{a}_n(\tilde{t}),r_n(\tilde{t}),\boldsymbol{s}_n(\tilde{t}+1))$ of size $\mathcal{T}$ is then sampled from the experience buffer and for each sample $\tilde{t}$ of the minibatch, the target $z_n(\tilde{t})$ is calculated based on the Q-learning principle as follows:
	\begin{align}
		z_n(\tilde{\tilde{t}}) = r_n(\tilde{t}) + \gamma Q_n'(\boldsymbol{s}_n(\tilde{t}+1),\mu'_n(\boldsymbol{s}_n(\tilde{t}+1)|\boldsymbol{\theta}_n ^{\mu'})|\boldsymbol{\theta}_n^{Q'}) 
		\label{eq10} 
	\end{align}
	
	The critic NN parameters $ \boldsymbol{\theta}_n^{Q} $ are then updated utilizing the gradient of the mean square error loss function defined as follows:
	\begin{align}
		\frac{1}{\mathcal{T}}\sum_{\tilde{t}}(z_n(\tilde{t}) - Q_n(\boldsymbol{s}_n(\tilde{t}),\boldsymbol{a}_n(\tilde{t})|\boldsymbol{\theta}_n^Q))^2
		\label{eq11} 
	\end{align}
	The actor NN weights $ \boldsymbol{\theta}_n^{\mathrm{\mu}} $ are then updated for each iteration by the sampled policy gradient to maximize the expected discounted reward obtained as follows \cite{DDPG}: 
	\begin{multline}
		\nabla_{\boldsymbol{\theta}_n^{\mu}} J(\boldsymbol{\theta}_n) \approx
        \\
        \frac{1}{\mathcal{T}}\sum_{\tilde{t}}\nabla_{\boldsymbol{a}_n}Q_n(\boldsymbol{s}_n(\tilde{t}),\boldsymbol{a}_n(\tilde{t})|\boldsymbol{\theta}_n^Q)\nabla_{\boldsymbol{\theta}_n^{\mu}} \mu_n(\boldsymbol{s}_n(\tilde{t})|\boldsymbol{\theta}_n ^{\mu} )
		\label{eq13} 
	\end{multline}	
	
	 The expression $ \nabla_{a_n}Q_n(\boldsymbol{s}_n(\tilde{t}),\boldsymbol{a}_n(\tilde{t})|\boldsymbol{\theta}_n^Q) $ is derived from the critic network, by backpropagating from its output with respect to the action input from $\mu_n(\boldsymbol{s}_n|\boldsymbol{\theta}_n ^{\mu})$. Two techniques have been utilized to enhance the stability of training for the DDPG actor-critic structure. These involve 1) utilizing an experience buffer $\boldsymbol{\mathrm{R}}_n$ for training the critic through  \eqref{eq11}, and 2) employing target networks for both the actor and critic that are updated using the Polyak averaging through \eqref{eq101}. The weights $\boldsymbol{\theta}_n ^{Q'}$ and $\boldsymbol{\theta}_n ^{\mu'}$ of the target networks are updated every $ t^p $ iteration using the Polyak averaging method with the target smoothing factor $\tau \in (0,\ 1]$, represented as follows: 
	\begin{align}
		&\boldsymbol{\theta}_n^{Q'}= (1-\tau)\boldsymbol{\theta}_n^{Q'} + \tau \boldsymbol{\theta}_n^Q \notag \\
		&\boldsymbol{\theta}_n^{\mu'}= (1-\tau)\boldsymbol{\theta}_n^{\mu'} + \tau \boldsymbol{\theta}_n^{\mu}
		\label{eq101} 
	\end{align} 
	
	Once the parameters of agent  $n$ are updated for a number of iterations, these parameters should be exchanged between the CPU and other agents in a federated framework. 
	The stated procedure for updating network parameters runs for $T$ iterations, after which the episode is increased by one unit. After each $T^{\mathrm{FL}}$ episode, each agent $n$ transmits its parameter vector $ \boldsymbol{\theta}_n $ to the CPU. The CPU then combines the agents’ parameter vectors to compute updated global NNs weights. There are several methods of accomplishing this \cite{FDRL1}, \cite{FDRL2}. A common approach is to simply average $ \boldsymbol{\theta}_n $ across all agents; this technique has been adopted in our work. The updated global NNs' weights are then sent back to the APs through fronthaul links in a federated scheme. 
	The objective of the FDRL algorithm is for a group of $ N $ agents to collectively learn a policy that performs optimally and consistently well throughout the entire environment. 
	Algorithm 1 outlines the FDRL algorithm for training agent $ n $. Furthermore,
	Fig. \ref{fig:FDRL} depicts the federated structure of the learning process employed in the cell-free RSMA network. Considering that the state vector $\boldsymbol{s}_n(t)$ and the action vector $\boldsymbol{a}_n(t)$ in \eqref{eq9911} involve fixed components $\boldsymbol{\tilde{s}}_{n'}(t_0)$, $\boldsymbol{\tilde{a}}_{n'}(t_0), \forall n'\neq n$ and dynamic components $\boldsymbol{\tilde{s}}_n(t)$, $\boldsymbol{\tilde{a}}_n(t)$, the corresponding NNs consist of two types of uncolored (white) and filled-colored circles. The rich-colored neurons correspond to the ones updated through the corresponding agent, and the faint-colored neurons indicate the ones affected through multiple agents due to the common UEs served by multiple APs in the cell-free framework.
	

	\begin{algorithm}[t]
		\caption{\small\!: FDRL algorithm for training agent $n$}
		\begin{algorithmic}[1]
			\Statex \hspace{-20pt} {\bf Inputs}: 
			\Statex $\bold{H}_{kn},\ \forall k\in \mathcal{K}_n, \boldsymbol{\theta}_n ^{Q}, \boldsymbol{\theta}_n ^{Q'}, \boldsymbol{\theta}_n ^{\mu}, \boldsymbol{\theta}_n ^{\mu'}$;
			\Statex \hspace{-20pt} {\bf Output}: 
			\Statex  $\bold{P}_{n}^{\mathrm{c}}, \bold{P}_{kn}^{\mathrm{p}}, c_k,\ \forall k\in \mathcal{K}_n$;
			\Statex Updated values of  $  \boldsymbol{\theta}_n ^{Q}, \boldsymbol{\theta}_n ^{Q'}, \boldsymbol{\theta}_n ^{\mu}, \boldsymbol{\theta}_n ^{\mu'}$;
			\Statex \hspace{-20pt} \textbf{Agent Parameters:}
			\Statex State and action ($\boldsymbol{s}_n$,$\boldsymbol{a}_n$) presented in \eqref{eq9911} and \eqref{eq9910};
			\Statex Reward function $r_n $: Refer to Algorithm 2;
			\Statex \hspace{-20pt} {\bf Main Procedure}
			\For{ episode $e =  1, ... ,E $}
			\For{iteration $ t = 1, ... ,T $}
			\State Select action $ \boldsymbol{a}_n(t) = \mu_n(\boldsymbol{s}_n(t)|\boldsymbol{\theta} ^{\mu})$, obtain reward $ r_n(t) $ from Algorithm 2, and observe new state $ \boldsymbol{s}_n(t+1) $;
			\State Store tuple $ (\boldsymbol{s}_n(t), \boldsymbol{a}_n(t), r_n(t), \boldsymbol{s}_n(t+1)) $ in the experience buffer $\bold{R}_n $;
			\State Sample a random mini-batch $ (\boldsymbol{s}_n(\tilde{t}),\boldsymbol{a}_n(\tilde{t}),r_n(\tilde{t}),\boldsymbol{s}_n(\tilde{t}+1)) $ from $  \bold{R}_n $; 
			\State Compute $z_n(\tilde{t})$ from \eqref{eq10};
			\State Update critic NN from \eqref{eq11};
			\State Update actor NN from \eqref{eq13};
			\If{$ \mathrm{mod}(t, t^p) = 0 $}
			\State Update target NNs from \eqref{eq101};
			\EndIf
			\EndFor
			\If{$ \mathrm{mod}(e, T^{\mathrm{FL}}) = 0 $}
			\State Transmit the agent's parameters to the CPU and receive all agents' combined parameters from the CPU in a federated scheme; 
			\EndIf
			\EndFor
		\end{algorithmic}
	\end{algorithm}	
	\begin{algorithm}[t]
		\caption{\small\!: Computation of the reward function for agent $n$}
		\begin{algorithmic}[1]
			\Statex \hspace{-20pt} {\bf Inputs}: 
			\Statex $\bold{H}_{kn}, \forall k \in \mathcal{K}_n$, action vector $\boldsymbol{a}_n$, episode $e$;
			\Statex \hspace{-20pt} {\bf Output}: 
			\Statex Reward value $r_n$;
			\Statex \hspace{-20pt} {\bf Main Procedure}: 
			\State Extract $\bold{P}_{kn'}^{\mathrm{p}},\bold{P}^{\mathrm{c}}_{n'},c_k ,\ \forall k, n'$ from $\boldsymbol{a}_n$;
			\If{$ \mathrm{mod}(e, T^{\mathrm{FL}}) = 0 $}
			\State Update $\tilde{\bold{H}}_{kn'}, \forall n'\neq n, \forall k$ from \eqref{eq92};
			\EndIf
			\State Compute $ \boldsymbol{\Sigma}_k^c $ and $ \boldsymbol{\Sigma}_k^p $ from \eqref{eq6} by letting $ \bold{H}_{kn'}= \tilde{\bold{H}}_{kn'},\ \forall n'\neq n, \forall k$ ;
			\State Compute $ R_k^{\mathrm{c}} $ and $ R_k^{\mathrm{p}} $ from \eqref{eq5}; 
			\State Compute $ R^{\mathrm{c}} $  from \eqref{eq7}; 
			\State Compute $ R_k, \ \forall k \in \mathcal{K}_n $ from \eqref{eq8};
			\State Calculate the reward output as $r_n= \mathrm{min}\  R_k, \forall k\in \mathcal{K}_n$;
		\end{algorithmic}
	\end{algorithm}
	
	\subsection{Reward computation}
	In this subsection, we propose the steps required for calculating reward value $r_n$. 
	The reward function plays a crucial role in the algorithm's performance, which we have considered here equal to the objective function of problem \eqref{eq91} evaluated only for $k\in\mathcal{K}_n$. More specifically, given that each agent has limited access to the information, instead of incorporating the objective function of \eqref{eq91} for all $k$, we evaluate it only for $k\in\mathcal{K}_n$. Regarding the RSMA formulations presented in Section II, the procedure required for calculating the reward value for each iteration is presented in Algorithm 2. We assume that for agent $n$, the CSI corresponding to $\bold{H}_{kn}, \forall k \in \mathcal{K}_n $ are known. Furthermore, we can estimate the CSI between each UE and other agents, denoted as $\tilde{\bold{H}}_{kn'},\ \forall k, n'\neq n$, based on the current action and state vector by utilizing \eqref{eq991}, as expressed in the following: 
	\begin{align}
		\tilde{\bold{H}}_{kn} = (\bold{P}_n^{\mathrm{c}}(\bold{P}_n^{\mathrm{c}})^{\mathrm{H}})^{-1} \bold{P}_n^{\mathrm{c}}
		(\mathrm{\hat{\bold{Y}}}_{kn}^{\mathrm{c}})^{\mathrm{H}}
		\label{eq92} 
	\end{align}
	
	It is important to note that $\bold{H}_{kn} $ is updated in each iteration of the training process, and $\tilde{\bold{H}}_{kn'}$ is updated in every $T^{\mathrm{FL}}$ episode. By utilizing $\bold{H}_{kn} $ and $\tilde{\bold{H}}_{kn'}$  as well as the values of the action vector, we can calculate $ \boldsymbol{\Sigma}_k^c $ and $ \boldsymbol{\Sigma}_k^p $ as specified in  \eqref{eq6}. Subsequently, we obtain the values of $ R_k^{\mathrm{c}} $, $ R_k^{\mathrm{p}} $ and $ R^{\mathrm{c}} $ from \eqref{eq5} and \eqref{eq7}.  Finally, from \eqref{eq8}, we compute $r_n$ as the minimum of $ R_k $ for $\forall k \in \mathcal{K}_n $.
	
	\subsection{Overall algorithm}
		
	\begin{algorithm}[t]
		\caption{\small\!: Overall training algorithm }
		\begin{algorithmic}[1]
			\Statex \hspace{-20pt} {\bf Output}: 
			\Statex $\boldsymbol{{\theta}}_n=\left\{ \boldsymbol{{\theta}}_n ^{Q}, \boldsymbol{{\theta}}_n ^{Q'}, \boldsymbol{{\theta}}_n ^{\mu}, \boldsymbol{{\theta}}_n ^{\mu'\!}\right\}\!,\boldsymbol{\tilde{\theta}}_n\!=\!\left\{ \boldsymbol{\tilde{\theta}}_n ^{Q}, \boldsymbol{\tilde{\theta}}_n ^{Q'}, \boldsymbol{\tilde{\theta}}_n ^{\mu}, \boldsymbol{\tilde{\theta}}_n ^{\mu'}\right\},
			\forall n$;
			\Statex \hspace{-20pt} {\bf Initialization:}
			\State Randomly initialize the critic $ Q_n(\boldsymbol{s}_n,\boldsymbol{a}_n|\boldsymbol{\theta}_n^Q) $ and the actor $ \mu_n(\boldsymbol{s}_n|\boldsymbol{\theta}_n ^{\mu})$
			with weights $ \boldsymbol{\theta}_n^Q $ and $ \boldsymbol{\theta}_n ^{\mu}$ respectively;
			\State Initialize the target networks $ Q' $ and $ \mu'$ with weights $ \boldsymbol{\theta}_n^{Q'} \leftarrow \boldsymbol{\theta}_n^Q$ and $ \boldsymbol{\theta}_n ^{\mu'} \leftarrow \boldsymbol{\theta}_n ^{\mu} $ ;
			\State  Initialize experience buffer $ \bold{R}_n = \O$ and $ \tilde{\bold{H}}_{kn'} = \bold{0}, \forall n'\neq n, k\in\mathcal{K}$;
			\Statex \hspace{-20pt} \textbf{Main Procedure:}
			\State \textbf{Do}
			\State \hspace{7pt} Initialize the environment according to the model of the system and observe the initial state vector $\boldsymbol{s}_n(1)$ by adopting new $\boldsymbol{\mathrm{H}}_{kn},\ \forall k\in\mathcal{K}$;
			\Statex {\bf Precoder Pre-training:} 
			\State \hspace{7pt} Set $g_{kn}=1,\ \forall n,k$;
			\State \hspace{10pt}Update  $  \boldsymbol{\theta}_n$ and obtain $ \bold{P}_{n}^{\mathrm{c}}, \bold{P}_{kn}^{\mathrm{p}}, c_k,\! \forall n,k$ using Alg. 1;
			\Statex {\bf AP Selection:} 
			\State \hspace{7pt} Update the AP selection parameters $g_{kn},\ \forall n,k$ calculated by the CPU through solving \eqref{eq96}; 
			\Statex {\bf Precoder Fine-tunning:} 
			\State \hspace{9pt} Obtain the precoders by repeating Step 7 for the second network $ \tilde{\boldsymbol{\theta}}_{ n } $, using the updated $g_{kn}$ and a smaller value for variance of exploration noise, where the initial value for $\tilde{\boldsymbol{\theta}}_{ n }$ is taken as  ${\boldsymbol{\theta}}_{ n }$ obtained from the pre-trained NN;%
			\State \textbf{While} Training is completed;
		\end{algorithmic}
	\end{algorithm}
	
\begin{figure}
		\centering
		\includegraphics [width=250pt,height=160pt]{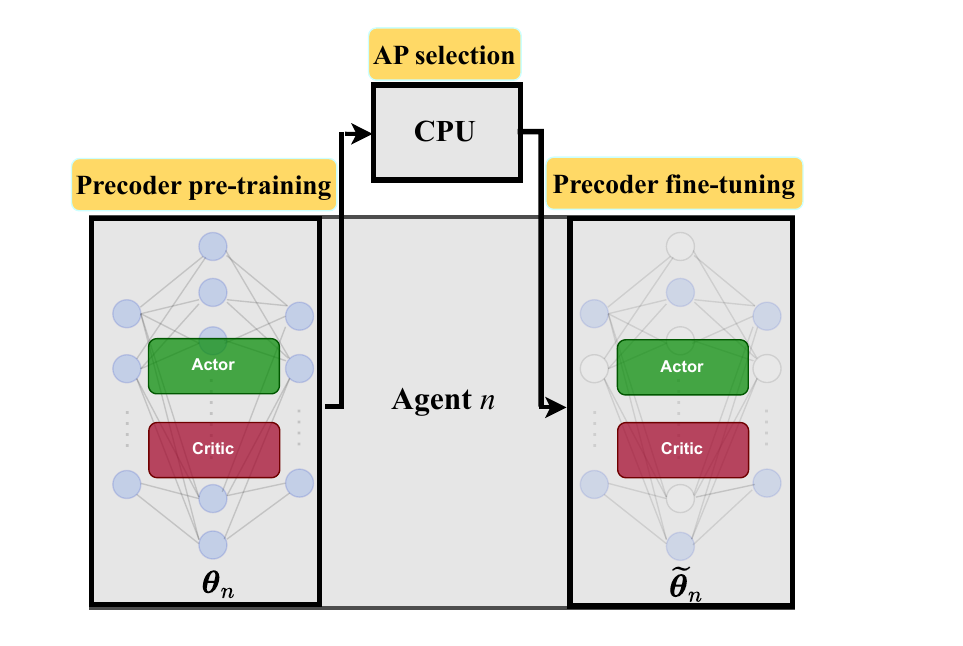} \\
		\caption{ A three-stage precoder design and AP selection procedure involved at the training phases of the proposed FDRL algorithm. While the FDRL precoder pre-training and fine-tunning are executed at the agent level, the AP selection algorithm is conducted at the CPU level.} 
		\label{fig:overall}
	\end{figure}

	The overall algorithm for solving Problem $\mathcal{P}1$, comprising precoding design and AP selection, is presented as Algorithm 3. 
    In the initialization phase, the actor and critic NNs and the experience buffer are initialized (Steps 1-3). 
	The training loop starts in Step 5 with initializing the environment and getting the channels according to the model explained in Section II.
	We employ a three-phase updating scheme corresponding to the three-block update process shown in Fig. \ref{fig:overall}. In the first block, which corresponds to Steps 6 and 7 of the algorithm, we perform a simple non-optimal AP assignment (e.g., $ g_{kn}=1,\ \forall k,n $) and then we address Problem $\mathcal{P}4$ 
	to achieve a {\bf precoder pre-training} using the proposed FDRL method. 
	In the second block of Fig. \ref{fig:overall}, corresponding to Step 8 of Algorithm 3, using the pre-trained precoders obtained from the previous step, we perform the {\bf AP selection} to get the values of $g_{kn},\ \forall n,k$ by solving \eqref{eq96}.	
	Finally, in the third block (Step 9 of Algorithm 3), we refine the precoders through another iteration of the FDRL algorithm. This time, we incorporate the updated AP selection parameters $g_{kn}$. To expedite convergence and reduce computational overhead, we employ transfer learning. By utilizing the policy trained in Step 7 as a starting point, we efficiently train more effective precoders in this final stage.
	Note that as illustrated in Fig. \ref{fig:overall}, the FDRL algorithm is executed at each agent, while the AP selection algorithm and the aggregation process of federated learning are handled at the CPU.   
	
	\begin{remark}	
		The design of learning-based precoding and AP selection involves two phases, including {\it training} and {\it inference}. Therefore, it is important to assess the computational and communication costs of each phase separately. Similar to most existing learning-based precoding designs outlined in the literature (e.g., [29] and [30]), the training phase of our proposed method is implemented offline. This offline training is generally performed during periods when the APs are not in use, ensuring that communication overhead does not affect system performance. Upon completion of model training, no additional communication expenses are incurred during the inference phase. Therefore, the computational complexity associated with the proposed algorithms does not raise concerns regarding their applicability.
	\end{remark}

	\subsection{Complexity analysis of proposed solution}
	In what follows we investigate the complexity of the algorithm relating to the training of each agent in the proposed federated learning-based framework. In the complexity analysis, we consider the structure employed in our numerical results, wherein each of the NNs of the agents consists of three layers. Let $n_0$, $n_1$, and $n_2$ denote respectively the number of neurons corresponding to the input, hidden, and output layer for a given actor/critic NN. In the feed-forward process, at each of the $n_1$ neurons of the hidden layer, the $n_0$ input values are multiplied by a weight, and then the weighted values are summed. The resulting sum then serves as the input of $\tanh(\cdot)$ activation function which is $\mathcal{O}(1)$ for each neuron.
	The neurons' outputs of the hidden layer are then weighted and summed, and then fed to the output layer. 
	Subsequently, this vector containing those sums is fed again into the $ \tanh(\cdot) $ activation function. Therefore, the feed-forward process complexity is obtained as $\mathcal{O}(n_1n_0 + n_1n_2)$.
	Similarly, the computational complexity of the back-propagation process involves computing the error signal, multiplying it by the corresponding weights of each neuron, and summing the results, which leads to equal computational complexity to that relating to the feed-forward process. For the actor/target-actor NNs we have $ n_0=2NK(M')^2$ and $ n_2=2NMM' + K$, and for the critic/target-critic NNs we have $n_0= 2NK(M')^2 + 2NMM' + K$ and $n_2=1$ respectively. Furthermore, we assume that $ n_1=\eta n_0$, where $\eta>1$ is a factor that represents how large is the size of the hidden layer compared to that of the input layer. $\eta$ is generally a constant whose value is independent of $n_0$ and $n_2$. Noting that the total computational complexity for a fully connected 3-layer NN through feed-forward and back-propagation is  $\mathcal{O}(n_1n_0 + n_1n_2)$, we conclude that the computational complexity at each training iteration for each of the actor, target-actor, critic, and target-critic NNs is obtained as  
    $\mathcal{O}(\eta N^2M'^2(K^2M'^2 + M^2+KMM'^2))$. For scenarios where the number of UEs is greater than the number of APs antennas, the complexity is reduced to $\mathcal{O}(\eta N^2M'^4K^2)$.
	
	\section{Numerical Results and Discussions}
	\label{sec:numerical}
	\begin{table}[htbp]
		\label{tb1}
		\caption{Simulation Parameters}
		\begin{center}
			\begin{tabular}{|l|l|l|}
				\hline
				\textbf{Parameter}&\textbf{Description}&{\textbf{Value}} \\
				\hline
				{$f$}&Carrier frequency&{1.9 GHz} \\
				\hline
				{$B$}&Bandwidth&{20 MHz} \\
				\hline
				{$N^{\mathrm{F}}$}&Noise figure&{9 dB} \\
				\hline
				{$h^{\mathrm{AP}}, h^{\mathrm{UE}}$}&Altitude of APs and UEs&{15 m, 1.65 m} \\
				\hline
				{$\overline{d},\underline{d}$}&Channel calculation parameters&{50 m, 10 m } \\
				\hline
				{$M$}&No. of APs antenna elements&{4} \\
				\hline
				{$M'$ (MIMO)}&No. of UEs antenna elements&2 \\
				\hline
				{$\mathcal{M}$, $\Omega$}& Nakagami distribution parameters&{1, 2} \\
				\hline
				{$ P^{\mathrm{max}} $}&Max. transmit power&{$30\ \mathrm{dBm W} $}\\
				\hline 
				{$ \overline{N}^{\mathrm{UE}} $}& Max. no. of serving APs for each UE&{$4$}\\
				\hline 
				{$ T^{\mathrm{FL}} $}&FDRL updating episode rate &{$ 10 $}\\
				\hline
			\end{tabular}
			\label{tab1}
		\end{center}
	\end{table}
	In this section, we provide simulation scenarios for analyzing the performance of the proposed FDRL-RSMA structure employed for cell-free MIMO networks. Consider a network area of $1 \mathrm{Km}\times 1\mathrm{Km}$ wherein the APs and UEs are randomly scattered with uniform distribution. The spacing between neighboring antenna elements is half a wavelength at all transmitters and receivers. 
	We consider that during each training step of Algorithm 3, the UEs' locations and consequently the environment are reset according to Step 5 of the algorithm. 
	In our work, we employ a discount factor of $ \gamma = 0.99 $, 
 constant parameter $\eta = 2$, and the Polyak averaging parameter of $\tau = 10^{-3}$. 
	The critic optimizer is Adam with its default hyper-parameters set as $\beta_1 = 0.9$, $\beta_2 = 0.999$, and a learning rate of $0.001$. Each agent's NNs structure comprises an input layer, a hidden layer, and an output layer, whose weights are initialized randomly. Additionally, we have employed $\tanh(\cdot)$ as the activation function. Unless explicitly stated, all other parameters' values are adopted from Table I.
	\begin{figure}[htbp]
			\includegraphics[height=4cm,width=8cm]{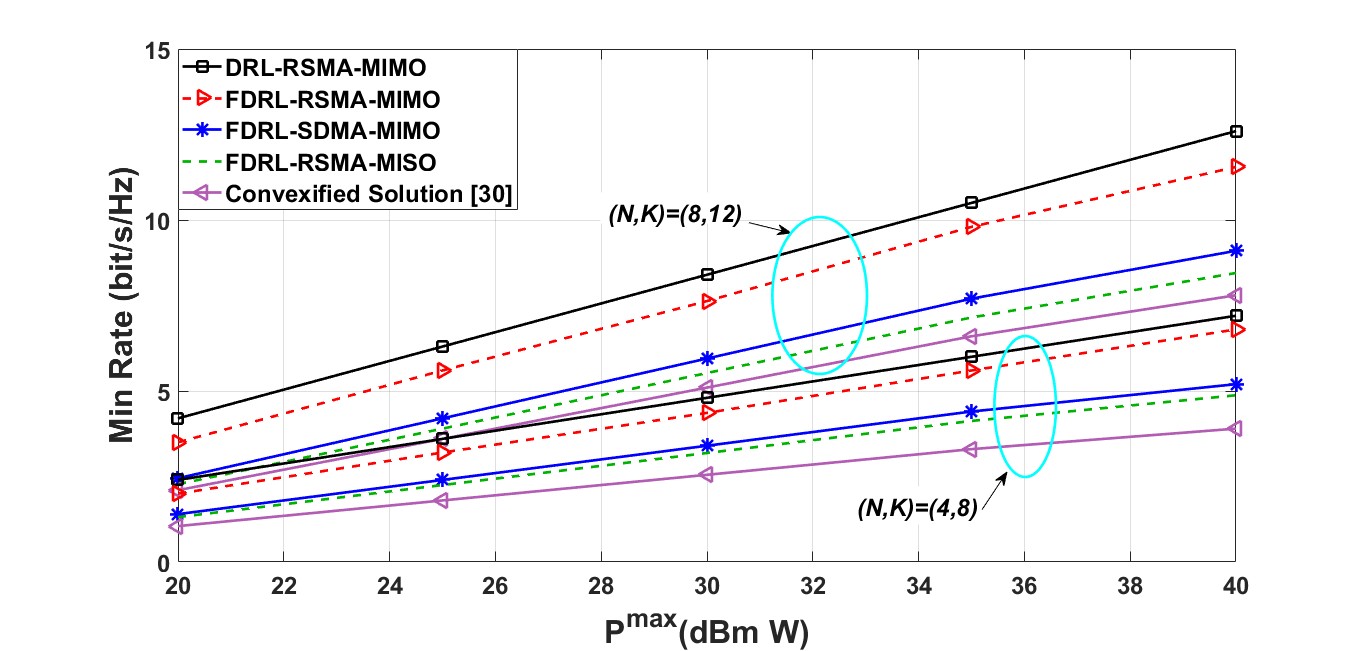}
			\hspace{-10pt}
		\caption{Minimum UE rate per $P^{\mathrm{max}}$ in various methods for $ (N,K)=(4,8) $, and $ (N,K)=(8,12) $.}
		\label{fig:R PER P}
	\end{figure}
	
	We consider various simulation methods employing different RSMA cell-free MIMO and MISO methods. These methods include the centralized DRL-RSMA-MIMO, the FDRL-RSMA-MIMO, the FDRL-RSMA-MISO, as well as the FDRL-SDMA-MIMO method\footnote{The SDMA method is a special case of RSMA wherein $\boldsymbol{\mathrm{P}}^{\mathrm{c}}_n=\boldsymbol{0}, \forall n$.}. 
In the centralized DRL method, the algorithm is completely run on the CPU, wherein all steps relating to the AP selection, precoder pre-training and precoder fine-tuning are carried out, with no exchange of learning data between the APs and the CPU.  As a result, this approach yields better performance compared to the FDRL method, but it compromises information security and increases network load.
	
	Fig. \ref{fig:R PER P} illustrates the minimum UE rate per $P^{\mathrm{max}}$ for the aforementioned methods. We have also compared these methods with the results obtained from the solution scheme proposed in \cite{convex}, wherein approximation schemes have been employed to transform the non-convex problem \eqref{eq9} into a convex problem. Two cases for the number of APs and UEs have been considered as $(N,K)=(4,8)$ and $(N,K)=(8,12)$.
	It is observed that the performance of all schemes increases by increasing  $P^{\mathrm{max}}$, and the proposed FDRL-RSMA-MIMO shows a small performance degradation to the optimal centralized DRL-RSMA-MIMO scheme. Besides, while both FDRL-RSMA-MISO and FDRL-SDMA-MIMO show nearly the same performance measure, they both outperform the convexified solution scheme proposed in \cite{convex}.

\begin{figure}[htbp]
		\begin{tabular}{c}
			\includegraphics[height=4cm,width=8cm]{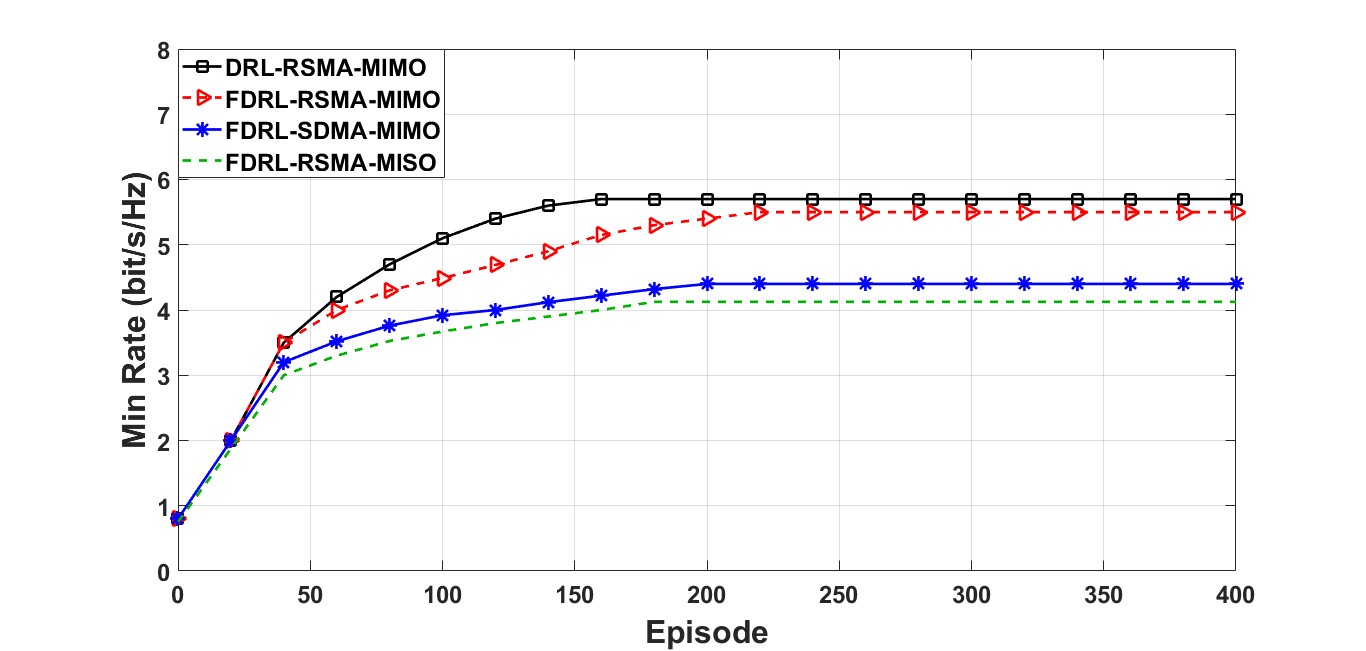}
			\\
			{\small(a)}
			\\
			\includegraphics[height=4cm,width=8cm]{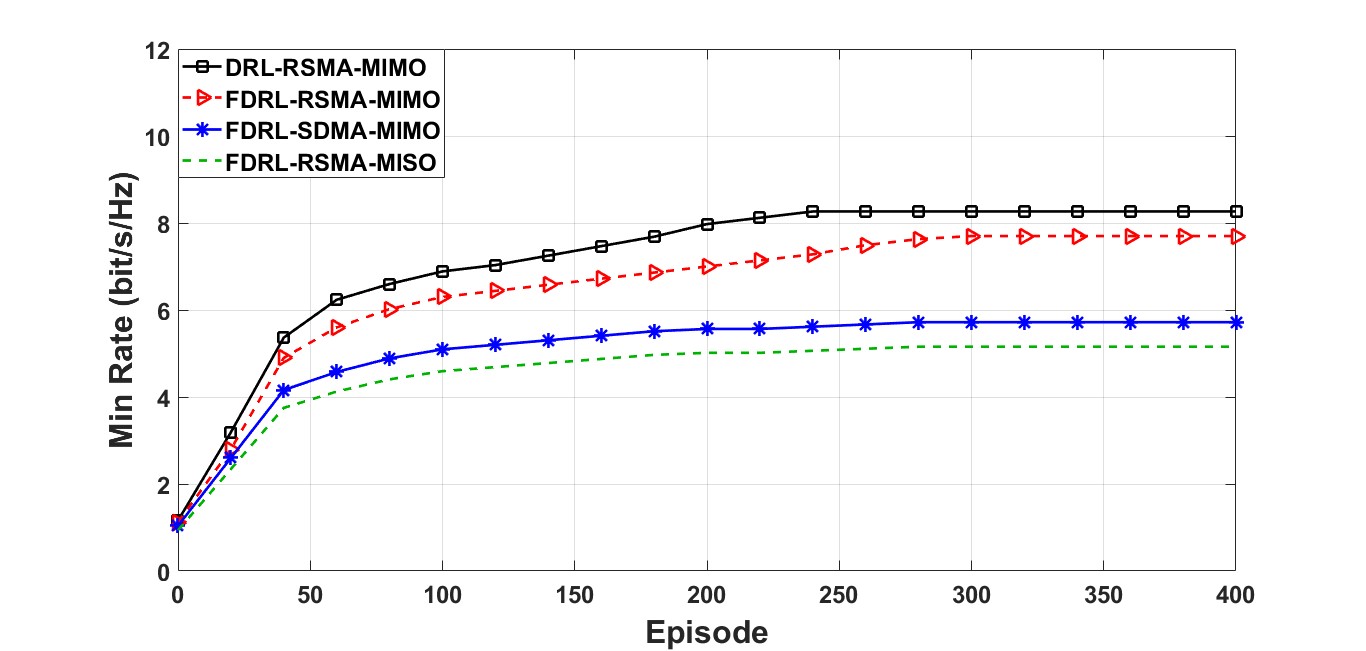}
			\\
			{\small(b)}
			\hspace{-10pt}
		\end{tabular}
		\caption{Minimum UE rate per episode in various methods for (a) $ (N,K)=(4,8) $, and (b) $ (N,K)=(8,12) $.}
		\label{fig:R per episod}
	\end{figure}

	
	Fig. \ref{fig:R per episod} illustrates the convergence of the FDRL and DRL methods for different multiple access and MIMO/MISO cases for $(N,K)=(4,8)$ and $(N,K)=(8,12)$. Firstly, it is seen that all schemes have converged after about 200 and 260 episodes in Figs. \ref{fig:R per episod}-(a) and \ref{fig:R per episod}-(b) respectively. Besides, considering the DRL-RSMA-MIMO and FDRL-RSMA-MIMO methods, while
    the DRL converges faster than the FDRL due to a central decision-making procedure, the final performance gap is seen to be marginal after convergence.  
    

\begin{figure}
		\centering
		\includegraphics [width=254pt]{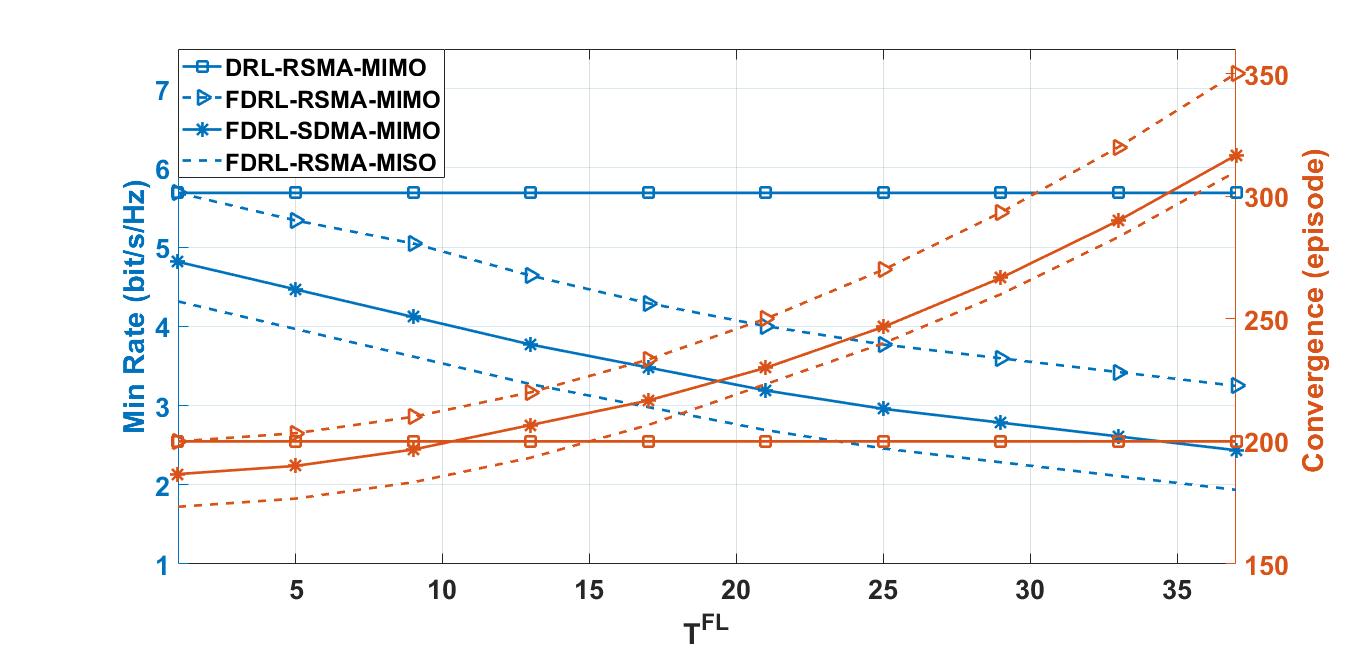} \\
		\caption{Minimum UE rate and convergence per $ T^{\mathrm{FL}} $ for $ (N,K)=(4,8) $.} 
		\label{fig:update}
	\end{figure}

	The rate of agents' information update ($1/T^{\mathrm{FL}}$) is an important factor that highly impacts the performance of federated learning schemes as illustrated in Fig. \ref{fig:update}. Note that in the centralized DRL method, we have a single agent where all parameters are updated at each episode, and thus the DRL performance is independent of $T^{\mathrm{FL}}$. It can be seen in the figure that a lower value of $T^{\mathrm{FL}}$ results in higher benefits in terms of both performance measures and convergence rates, at the cost of increased signaling overload. Besides, for the case where all FDRL agents are updated at each episode ($T^{\mathrm{FL}}=1$), the FDRL shows the same performance as the DRL in a more secure way by preserving the privacy of the agents' information. Moreover, as illustrated in Fig. \ref{fig:update}, the FDRL-SDMA-MIMO and FDRL-RSMA-MISO methods demonstrate faster convergence compared to FDRL-RSMA-MIMO. This increased speed of convergence is attributed to the use of lower-dimensional NNs in these methods.
	
	
\begin{figure}
		\centering
		\includegraphics [width=254pt]{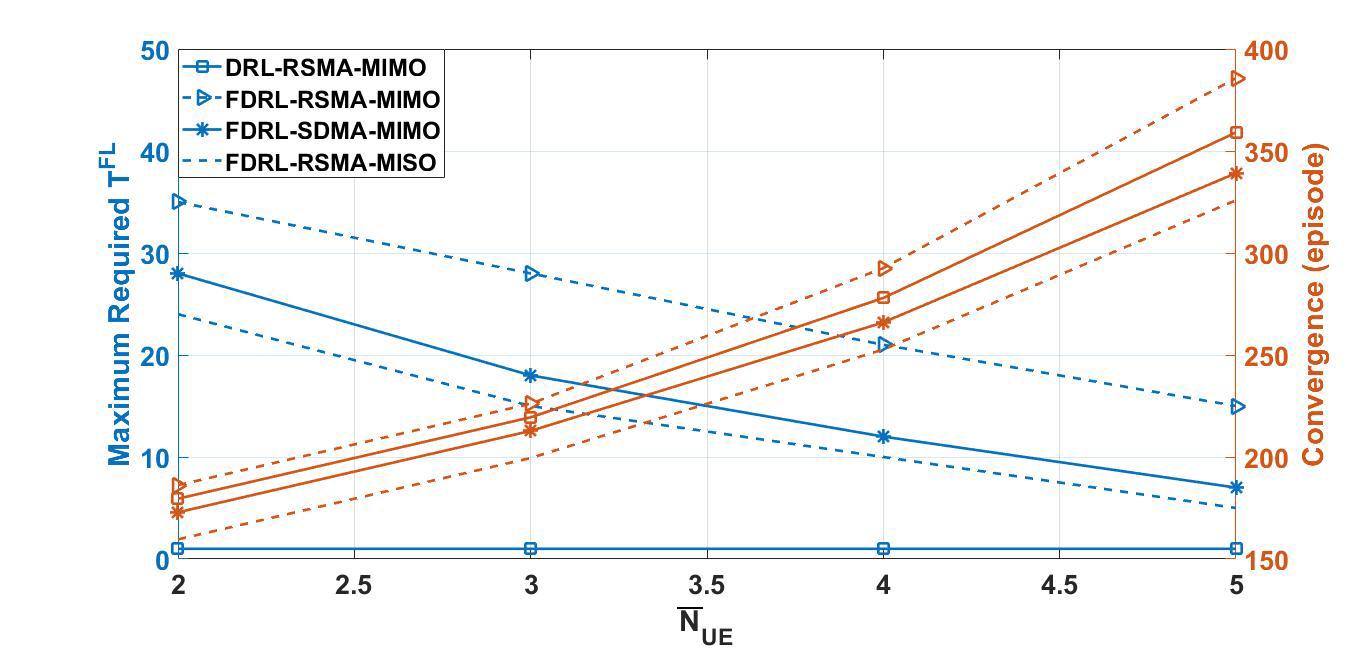} \\
		\caption{Maximum required $ T^{\mathrm{FL}} $ as well as convergence per $\overline{N}^{\mathrm{UE}}$ to achieve $\mathrm{Min}(R_k) =5$ bit/s/Hz for $ (N,K)=(8,12) $.} 
		\label{fig:common}
	\end{figure}

	The maximum number of cell-free APs serving a single UE, denoted by $\overline{N}^{\mathrm{UE}}$, can highly impact the performance of the proposed FDRL-based resource management scheme as illustrated in Fig. \ref{fig:common}. This figure shows the minimum required $ T^{\mathrm{FL}} $ as well as the convergence per $\overline{N}^{\mathrm{UE}}$ to achieve a given performance measure $\mathrm{Min}(R_k) =5$ bit/s/Hz.
	The parameter $\overline{N}^{\mathrm{UE}}$ is an indicator showing how effectively the network leverages the distributed cell-free network setup. A unity value corresponds to a pure conventional cellular network, while higher values incorporate more cell-free intrinsic properties. As observed, lower values of $\overline{N}^{\mathrm{UE}}$ result in a lower gap in convergence speed between the performance of FDRL-RSMA-MIMO and DRL-RSMA-MIMO algorithms.    
	This relates to the higher independence of the agents, which reduces the requirement of sharing the information between the agents and the CPU. Moreover, higher values of $\overline{N}^{\mathrm{UE}}$ result in an increasing rate of information update ($1/T^{\mathrm{FL}}$) to achieve the given minimum UE rate. 
	
	\section{CONCLUSIONS}
	\label{sec:conclusions}
	In this paper, we presented a model for cell-free MIMO networks incorporating RSMA in the downlink mode. We evaluated the performance of the proposed model by deriving transmission rates for both the common and private components of RSMA. Additionally, we formulated an optimization problem to determine the optimal RSMA precoders and AP selection, aimed at maximizing the minimum rate of UEs. To address this problem, we proposed a three-stage algorithm that integrates AP selection using the PCA method and employs RSMA precoder pre-training and fine-tuning within a federated architecture as a distributed machine learning optimizer. Our results demonstrate that the proposed solution achieves performance comparable to the centralized benchmark while enhancing security and reducing the processing load at the CPU.


\end{document}